\documentclass[useAMS,usenatbib]{mn2e}

\usepackage[pdftex]{graphicx}
\usepackage{color}

\newcommand{\ch}{${\rm R_h}$}
\newcommand{\edit}{}

\title[The Cosmic Horizon is still no horizon!]{Phantom Energy and the Cosmic Horizon: \\
\ch\ is still not a horizon!}
\author[Geraint F. Lewis]{Geraint F. Lewis\thanks{E-mail:
geraint.lewis@sydney.edu.au}\\
Sydney Institute for Astronomy, School of Physics, A28, The University of Sydney, NSW 2006, Australia
}
\begin{document}

\date{\today}

\pagerange{\pageref{firstpage}--\pageref{lastpage}} \pubyear{2002}

\maketitle

\label{firstpage}

\begin{abstract}
There has been a recent spate of papers on the Cosmic Horizon, an apparently fundamental, although 
unrecognised, property of the universe. The misunderstanding of this horizon, it is claimed, demonstrates that
our determination of the cosmological makeup of the universe is incorrect, although several papers
have pointed out key flaws in these arguments. Here, we identify additional flaws in the most recent claims of
the properties of the Cosmic Horizon in the presence of phantom energy, simply demonstrating that it  
does not act as a horizon, and that its limiting of our view of the universe is  a trivial statement.
\end{abstract}


\section{Introduction}\label{introduction}
The presence of various horizons within our cosmological models have greatly elucidated 
our understanding of the workings of the universe, with both the particle and event horizons 
limiting the connexions between past and future cosmological events \citep{1956MNRAS.116..662R}. 
The universe also possesses a Hubble Sphere, which is not  a horizon, and is the distance from an observer 
that comoving objects are moving, due to the cosmic expansion, at the speed of light 
\citep[in proper coordinates see][]{1991ApJ...383...60H}. 

There has been the claim that the universe possess another, previously 
unidentified, horizon, dubbed the Cosmic Horizon (\ch), and the presence of this 
horizon significantly influences our observations of the 
cosmos~\citep{2007MNRAS.382.1917M,2009IJMPD..18.1113M,2009IJMPD..18.1889M,2012MNRAS.419.2579M,2012MNRAS.421.3356B,2012AJ....144..110M},
although several authors have demonstrated that a number of the key claims about the Cosmic
Horizon are incorrect~\citep{2010MNRAS.404.1633V,2012MNRAS.423L..26L,2012MNRAS.425.1664B}.

In this brief contribution, we discuss the recent claims of \ch\ in the presence of phantom energy \citep{2012JCAP...09..029M},
again showing them to be demonstrably incorrect. In Section~\ref{cosmichorizon} we briefly review the nature of 
\ch, while in Section~\ref{photon} we discuss what \ch\ means for the path of a photon traveling 
though an expanding universe.
Section~\ref{evolvinghorizons} demonstrates that \ch\ still fails to behave as an
unrecognised horizon in limiting our view of the universe, and the conclusions are presented in
Section~\ref{conclusions}. 

\section{The Cosmic Horizon}\label{cosmichorizon}
The concept of the Cosmic Horizon, \ch, was introduced by \citet{2007MNRAS.382.1917M} who, in rewriting the 
standard Friedmann-Robertson-Walker (FRW) invariant interval in `observer-dependent coordinates' found metric
terms that appeared to be singular at a proper distance of \ch=$1/H$, where $H$ is the Hubble constant; it is important
to remember that $H$ evolves over cosmic time, and so \ch\ is a similarly evolving proper distance from an observer.
In this initial work, \citet{2007MNRAS.382.1917M} claimed that the divergence of the metric components showed that 
\ch\ represented an infinite redshift surface, and hence a limit to our view of the universe. However, this was shown
as being  incorrect by \citet{2010MNRAS.404.1633V}, who demonstrated that the infinite redshift was due to a unphysical
choice for the coordinate velocity of an emitter at a distance of \ch, and correctly accounting for the coordinate transformation
between FRW and observer coordinates results in the same redshift; hence we can see photons that have traveled through
\ch.

As noted in  \citet{2010MNRAS.404.1633V}, the un-horizon-like properties come as no surprise as \ch\ is exactly the same  as the Hubble 
Sphere, a very well understood concept in cosmology \citep{1991ApJ...383...60H}. While in our current cosmology the 
Hubble Sphere will eventually become coincident with our event horizon \citep[see Figure 1 of][]{2004PASA...21...97D}, 
it is not, in itself, a horizon. 

However, there have been continuing claims about the fundamental nature of 
\ch, and recently, \citet{2012MNRAS.421.3356B} considered the paths of photons over cosmic history, examining
the proper distance traversed since the Big Bang. In considering several standard cosmological models, they concluded 
that the fundamental property of \ch\ is that any photon we receive today cannot have traveled from a distance greater than
\ch\ today. However, this was also 
shown to be incorrect by \citet{2012MNRAS.423L..26L} who  demonstrated that \ch\ does not have to
continuously grow or asymptote to a particular distance, and that if dark energy is actually 
of the form of phantom energy (with an equation of state of $\omega <  -1$), then \ch\ can decrease. Hence, light rays 
arriving at an observer can have travelled from a larger proper distance than \ch\ today
\citep[see Figure 3 of][]{2012MNRAS.423L..26L}.

The question of the influence of phantom energy on \ch\ was revisited in \citet{2012JCAP...09..029M}, who again 
redefined the cosmological properties of this Cosmic Horizon, concluding two key features of the photon paths should
reassure us of its fundamental importance. These are that;
\begin{quote}
``The most important feature of these curves is that none of those actually reaching us [today] ever attain 
a proper distance greater than the maximum extent of our cosmic horizon."
\end{quote}
and
\begin{quote}
``every null geodesic that possesses a second turning point ...., diverges to infinity"
\end{quote}
In the remainder of this paper, we will demonstrate that the first assertion is trivial, and the second is incorrect.

\section{To a photon, Just what is \ch?}\label{photon}
As we have noted previously, \ch\ simply corresponds to the Hubble Sphere 
\citep{2010MNRAS.404.1633V,2012MNRAS.423L..26L}, the distance from an observer at which the universal expansion
results in a proper velocity of the speed of light for a comoving object. In this section, we will discuss the what passing through the 
Hubble Sphere means to a photon, although it should be noted that this has been discussed in detail previously
\citep[e.g.][]{1993AmJPh..61..883E}.

Starting with the FRW invariant interval for a spatially flat expanding space-time, 
\begin{equation}
ds^2 = - dt^2 + a(t)^2 \left( dr^2 + r^2 d\Omega^2 \right)
\label{FRW}
\end{equation}
where $r$ is the comoving radial coordinate, $a(t)$ is the time-dependent scale factor, and 
$d\Omega$ considers the angular coordinates. With this, the proper distance to an object at a 
comoving distance of $r$ is $d = a(t)\ r$. For an observer at rest with regard to the comoving
spatial coordinates, the experienced proper time is the same as the cosmic time, $t$, and so we can
talk of the relative velocity of the distant object with regards to the observer as being
\begin{equation}
\dot{d} = \dot{a} r + a \dot{r}
\label{velocity}
\end{equation} 
where the dots denote derivatives with regards to $t$, and the $\dot{r}$ represent motion relative to the 
comoving coordinates. 

\begin{figure}
\includegraphics[scale=0.46, angle=0]{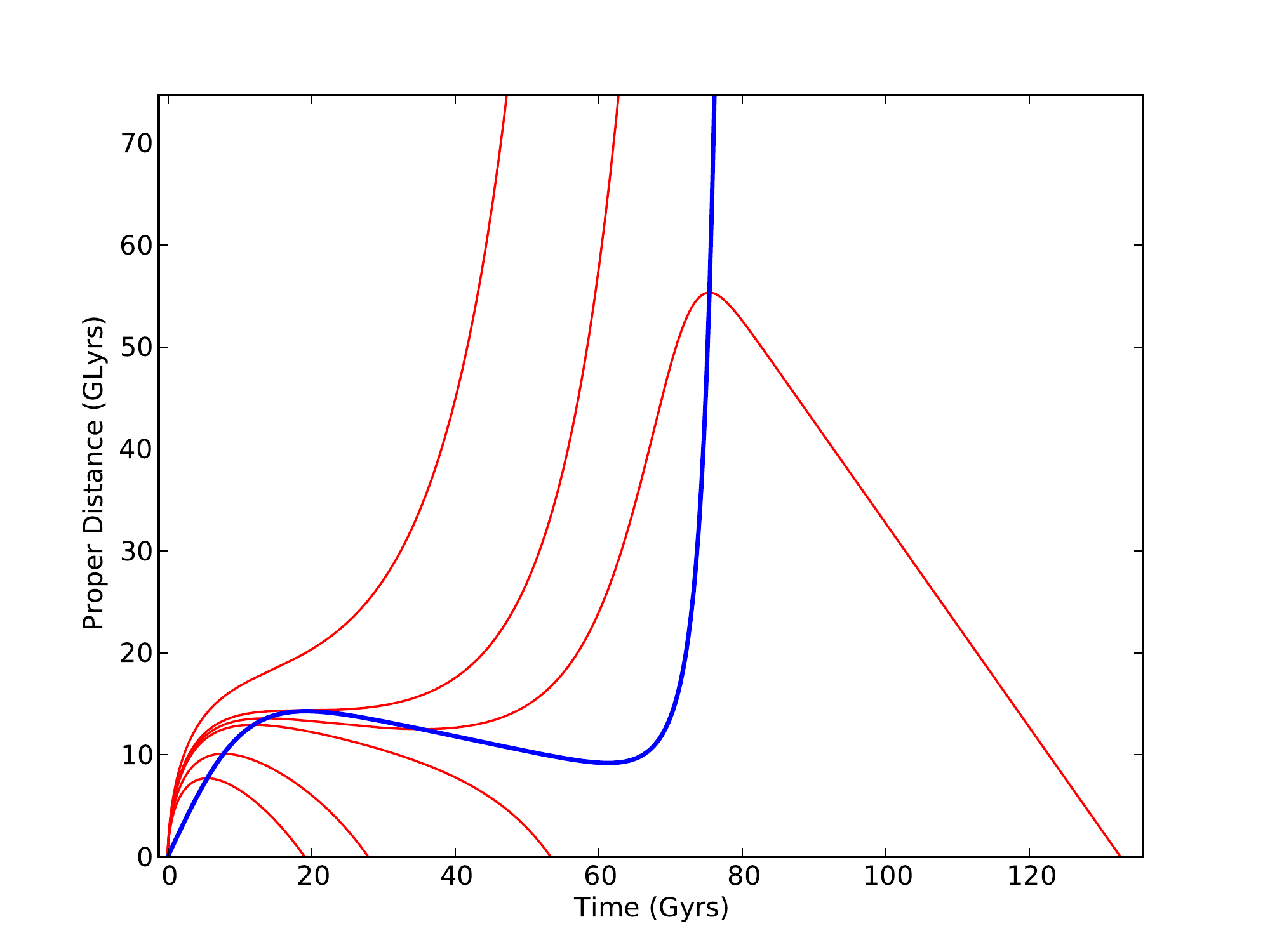}
\caption{The blue curve presents \ch\ for the cosmological model outlined in Section~\ref{evolvinghorizons},
while the red lines present a series of light rays emanating from the Big Bang (at the origin) and into the future.
Note that the abscissa represent cosmic time, whereas the ordinate is proper distance from an observer (at a
proper distance of zero). In this representation, four light rays arrive back at the observer; the key light ray is
the one arriving at $\sim$130 Gyrs, as this has crossed \ch\ three times.
\label{figure1}
}
\end{figure}

If we consider a photon moving in the radial direction, so the angular terms in Equation~\ref{FRW} can 
be neglected, and remembering that a photon path follows $ds = 0$, then it is straight-forward to show
\begin{equation}
\dot{r} = \pm\frac{1}{a}
\label{photongradient}
\end{equation}
and substituting the negative solution, as we are interested in a photon approaching the observer, 
into Equation\ref{velocity}, we find
\begin{equation}
\dot{d} = \frac{\dot{a}}{a} \left( a r \right) - a \left( \frac{1}{a} \right) = H d - 1
\label{velocity2}
\end{equation}
so at proper distances of $d > 0$, the photons are traveling, relative to the observer, at velocities not equal to
 the speed of light. At the distance of $d = 1 / H$,  which is \ch\ (or, in its proper parlance, the Hubble Sphere), 
then it is simple to see that $\dot{d} = 0$, and the photon is at rest with regards to the observer (in proper 
coordinates).

What this means is that, in proper coordinates, a photon crossing \ch\ represents an extremal or inflection point in the photon's 
path, and, considering that the photon path began at $r=0$ in the Big Bang, and returns to $r=0$ at a later time (i.e. 
it is detected by an observer), there must be a `most-distant' turn-around point in the photon's journey, where it stops
heading outward and starts heading inwards, and this must coincide with the photon crossing \ch.

In summary, the claim by \citet{2012JCAP...09..029M} that \ch\ represents a bound to the observed photon's 
path is nothing more than saying ``The maximal distance from which we receive a photon is no more than the 
largest distance at which it turns around in it journey"; this is a trivial statement.

\section{Evolving Horizons}\label{evolvinghorizons}
As a demonstration of the meaning of the Cosmic Horizon, we consider a universe in which \ch\ can be tuned to be
increasing or decreasing by modulating the equation of state of the dark energy component.
We begin by adopting the cosmological parameters of $\Omega_M=0.27$, $\Omega_\omega=0.73$ 
and $H_o = 72\ km/s/Mpc$. Unlike standard cosmological models, we allow the equation of state of the dark
energy component, $\omega$, to change as a function of time, adopting an evolutionary form given by
\begin{equation}
\omega(t) = -1.1 + \frac{1.43}{1 + exp\left( \frac{ 6 - t_h  }{ 0.3 } \right) }
\label{equationofstate}
\end{equation} 
where $t_h = t / ( 13.58\ GYrs )$ is the scaled cosmic time; it should be noted that this evolution is not physically motivated
and simply provides a model in which we can examine the corresponding evolution of \ch. {\edit With this form, the transition
in the form of the dark energy component takes place at $t \sim 6 t_h$, and the duration of the change over is $\sim 0.3 t_h$, 
although the values were chosen for purely illustrative purposes.
Essentially, in the early universe,
the dark energy component has an equation of state of $\omega\sim-1.1$, corresponding to a phantom energy, 
while around $t \sim 5 t_h \sim 65 Gyrs$ the
equation of state begins to transition to $\omega\sim\frac{1}{3}$, representing the equation of state of a relativistic mass-less
fluid, such as photons \citep[see][]{1988A&A...206..175L}. Note that given this evolution, expansion means that the universe
will become matter dominated in the future, given that the energy density in the now photon-like dark energy component
diminishes faster with cosmological expansion.}

In Figure~\ref{figure1}, we present the key properties of this universe in terms of the cosmic time (abscissa) and 
proper distance (ordinate). The blue curve denotes the evolution
of \ch, whereas the red curves represent light rays which emanate from the Big Bang (at the origin) and travel 
into the future. The left-hand half of the plot can be compared directly with Figure 3 of 
\citet{2012MNRAS.423L..26L}, with \ch\ increasing when the universe is matter dominated, and then decreasing 
as the phantom energy comes to dominate.  However, as the equation of state of the dark energy component 
evolves towards $\omega\sim\frac{1}{3}$, then \ch\ rapidly increases.

{\edit
An examination of the light rays in Figure~\ref{figure1} show that, in this representation of the universe, all photons head 
outwards after the Big Bang.  
Following this initial motion, three of the photon paths then encounter the evolving \ch\ only once, and then arrive back 
at the observer; this can be simply understood as, in proper 
coordinates, \ch\ marks the turning point in a photon's path. 
One photon path in Figure~\ref{figure1}
 does not encounter \ch, so that  there is no turning point in its path and  it does not return to the observer. 

There are two additional light rays which cross \ch\ more than once. Again, these 
rays are initially heading away from the origin
after the Big Bang, and both of them cross \ch\ and begin to head back towards the observer. 
However, due to the presence of the phantom energy component,  both photon paths 
encounter a decreasing \ch\ and the motion is reversed and the paths again move away from the observer; note that 
while one of these paths appear have a point of inflection, just touching \ch, it actually does possess two crossings.
As the equation of state of the dark energy component is changing and influencing the cosmic expansion, while one
photon path escapes to larger distance, the
other  is again reversed, at a point where the photon passes through  \ch\, and it now continues its journey towards the
observer, arriving at a cosmic time of $t \sim 130\ GYrs$. This is in stark contrast to the claim made by \citet{2012JCAP...09..029M}. 
}

{\edit
It is important to note that the results presented in this paper do not depend upon the specific form of the evolving 
dark energy described by Equation~\ref{equationofstate}. If we consider a universe with single energy component
described by an equation of state $\omega$, then \ch\ evolves as
\begin{equation}
\dot{R}_h = \frac{3}{2} \left( 1 + \omega \right)
\label{evolving}
\end{equation}
and hence for phantom energies, with $\omega<-1$, \ch\ would decrease, for a cosmological constant $(\omega=-1)$ 
it is a constant, and all other fluids, with $\omega>-1$, \ch\ increases. If Equation~\ref{equationofstate} is modified so that
the ultimate equation of state of the dark energy component  is $\omega=0$, corresponding to matter, \ch\ will increase in the future, 
similar to the evolution shown in Figure~\ref{figure1}. Again outward moving light rays encountering this increasing
\ch\ will change direction and will head back to the observer.
}

An examination of the discussion presented here should convince the reader
that by modifying the equations of state of the energy components of the universe, we could ensure that \ch\
oscillates through a arbitrarily complex path, and similarly that photon journeys can be made to arbitrarily change their
direction of motion toward or away from an observer. Each change of direction is accompanied with the photon
crossing in and out of \ch, which is extremely un-horizon-like behaviour (but precisely what you would expect as \ch\
being a turning point in a photon's path).

\section{Conclusions}\label{conclusions}
We have examined the most recent claims of the fundamental nature of the Cosmic Horizon, \ch, made by
\citet{2012JCAP...09..029M}, especially its behaviour in the presence of phantom energy. We have 
demonstrated that, by modifying the equation of state of the energy components in the universe, then \ch\ 
can be made to grown and shrink arbitrarily. Furthermore, as \ch\ represents the location of where a photon
changes direction, or goes through an inflection point, (in proper coordinates) relative to an observer, an
oscillating photon path can cross \ch\ a multitude of times before arriving at an observer. The 
'fundamental' property of \ch\ in terms of the farthest proper distance from which an observer receives a
photon is essentially a trivial statement.

As we have stressed in other contributions \citep[e.g.][]{2012MNRAS.423L..26L}, the importance and evolution of
true cosmic horizons is well understood, as is the mean of the Hubble Sphere~\citep{1991ApJ...383...60H}. 
Other than the trivial, the Cosmic 
Horizon of \citet{2012JCAP...09..029M} does not present a fundamental limit to our view of the universe.

\section*{Acknowledgments}
Luke Barnes and Krzysztof Bolejko are thanked for interesting discussions, and Brendon Brewer
is thanked for his insights into Python. 



\begin{thebibliography}{99}
%
\bibitem[\protect\citeauthoryear{Bikwa, Melia, 
\& Shevchuk}{2012}]{2012MNRAS.421.3356B} Bikwa O., Melia F., Shevchuk A., 2012, MNRAS, 421, 3356
%
\bibitem[\protect\citeauthoryear{Bilicki 
\& Seikel}{2012}]{2012MNRAS.425.1664B} Bilicki M., Seikel M., 2012, MNRAS, 425, 1664 
%
\bibitem[\protect\citeauthoryear{Davis 
\& Lineweaver}{2004}]{2004PASA...21...97D} Davis T.~M., Lineweaver C.~H., 2004, PASA, 21, 97
%
\bibitem[\protect\citeauthoryear{Ellis 
\& Rothman}{1993}]{1993AmJPh..61..883E} Ellis G.~F.~R., Rothman T., 1993, AmJPh, 61, 883 
%
\bibitem[\protect\citeauthoryear{Harrison}{1991}]{1991ApJ...383...60H} 
Harrison E., 1991, ApJ, 383, 60
%
\bibitem[\protect\citeauthoryear{Lewis 
\& van Oirschot}{2012}]{2012MNRAS.423L..26L} Lewis G.~F., van Oirschot P., 2012, MNRAS, 423, L26 
%
\bibitem[\protect\citeauthoryear{Linder}{1988}]{1988A&A...206..175L} Linder E.~V., 1988, A\&A, 206, 175 
%
\bibitem[\protect\citeauthoryear{Melia}{2007}]{2007MNRAS.382.1917M} Melia 
F., 2007, MNRAS, 382, 1917 
%
\bibitem[\protect\citeauthoryear{Melia}{2009}]{2009IJMPD..18.1113M} Melia 
F., 2009, IJMPD, 18, 1113 
%
\bibitem[\protect\citeauthoryear{Melia 
\& Abdelqader}{2009}]{2009IJMPD..18.1889M} Melia F., Abdelqader M., 2009, IJMPD, 18, 1889 
%
\bibitem[\protect\citeauthoryear{Melia 
\& Shevchuk}{2012}]{2012MNRAS.419.2579M} Melia F., Shevchuk A.~S.~H., 2012, MNRAS, 419, 2579 
%
\bibitem[\protect\citeauthoryear{Melia}{2012a}]{2012JCAP...09..029M} Melia 
F., 2012a, JCAP, 9, 29 
%
\bibitem[\protect\citeauthoryear{Melia}{2012b}]{2012AJ....144..110M} Melia 
F., 2012b, AJ, 144, 110 
%
\bibitem[Rindler(1956)]{1956MNRAS.116..662R} Rindler, W.\ 1956, MNRAS, 
116, 662 
%
\bibitem[\protect\citeauthoryear{van Oirschot, Kwan, 
\& Lewis}{2010}]{2010MNRAS.404.1633V} van Oirschot P., Kwan J., Lewis G.~F., 2010, MNRAS, 404, 1633 
%
\end{thebibliography}
\end{document}